# Role of Momentum Interpolation Mechanism of the Roe Scheme in Shock Instability

Xiao-dong Ren[1,2]    Chun-wei Gu[1]    Xue-song Li[1,*]

1. *Key Laboratory for Thermal Science and Power Engineering of Ministry of Education, Department of Thermal Engineering, Tsinghua University, Beijing 100084, PR China*

2. *Department of Mathematics, School of Science, Hong Kong University of Science and Technology, Hong Kong, China*

**Abstract:** The shock instability phenomenon is a famous problem for the shock-capturing scheme. By subdividing the numerical dissipation of the Roe scheme, the pressure-difference-driven modification for the cell face velocity is regarded as a version of the momentum interpolation method (MIM). The MIM is necessary for low Mach number flows to suppress the pressure checkerboard problem. Through the analysis and numerical tests, MIM has been discovered to have the most important function in shock instability. In fact, MIM should be completely removed for non-linear flows. However, unexpected MIM is activated on the cell face nearly parallel to the flow for high Mach number flows or low Mach number flows in shock. Therefore, the MIM should be kept for low Mach number flows and be completely removed for high Mach number flows and low Mach number flows in shock. For such conditions, two coefficients are designed on the basis of the local Mach number and a shock detector. The improved Roe scheme is then proposed. This scheme considers the requirement of MIM for incompressible and compressible flows and is validated for good performance of numerical tests. It is also proved that the acceptable result can be obtained with only Mach number coefficient for general practical computation. Therefore, the aim of decreasing rather than increasing numerical dissipation to cure shock instability can be achieved with very simple modification. More important, the mechanism of the shock instability has been deeply understood, where the MIM plays the most important role although it is not the only factor.

---

\* Corresponding author. Tel.: 0086-10-62794617; fax: 0086-10-62795946
E-mail address: xs-li@mail.tsinghua.edu.cn (X.-S. Li).

## 1. Introduction

The Roe scheme [1], which is one of the most important shock-capturing schemes, has obtained great success for simulating compressible flows and automatic capturing shock over the past decades. However, the Roe scheme also faces disastrous failings for certain problems, such as low Mach number incompressible flows [2–4] and the shock instability phenomenon of hypersonic flow computation.

Shock instability has different forms, such as the famous carbuncle phenomenon for supersonic flows around a blunt body, the kinked Mach stem for a double-Mach reflection flow, and the odd–even decoupling for a planar moving shock. Quirk [5] found that shock instability has various forms and proposed a curing method. These different forms have the same inherent mechanism, and a method that fails or succeeds in one form will fail or succeed in other forms.

The curing method for shock instability can be generally categorized into three groups. The first group adds dissipation to the scheme. More dissipation helps damp out spurious oscillation. Quirk [5] suggested combining a dissipative scheme, such as HLLE, and a less dissipative and more accurate scheme, such as Roe, through a switch sensor. Entropy fix is another classical method of adding extra dissipation by limiting the minimum system eigenvalue, and it has many improved versions [6, 7]. Qu and Yan [8] proposed the RoeMAS scheme that is effective for curing the shock instability by increasing basic upwind dissipation. The second group considers that the shock instability is due to the contradiction between the multi-dimensional characteristic of the compressible flow and the one-dimensional grid-aligned nature of the scheme. Therefore, multi-dimensional Riemann solvers have been developed, such as the rotated Roe-type schemes [9, 10]. The last group explains that the reason for shock instability is the pressure difference in the numerical mass flux [11]. Kim [12] also analyzed the effect of the pressure difference through a linear perturbation analysis and then proposed an improved Roe version called the RoeM scheme.

Extending the shock-capturing scheme to all-speed flows is another important subject for all schemes such as Roe-type [2–4], HLL-type [13, 14], and AUSM-type [15, 16] schemes. While studying an all-speed Roe scheme, Ref. [17] discovered that the Roe scheme has an inherent mechanism to prevent the checkerboard problem, which is a classical problem in the computation of incompressible flows. The checkerboard problem refers to the chess-like pressure jump due to the pressure–velocity decoupling. The inherent mechanism is identified as the pressure-difference-driven modification for the numerical cell face velocity, which is the only term discussed in Ref. [11, 12]. Moreover, it can be regarded as a version of the momentum interpolation method (MIM) [18], which is the classical method to suppress the pressure checkerboard and has many forms [19–21]. Ref. [4] further discussed the mechanism and showed that the different versions of MIM can replace each other in a pressure-difference modification for cell face velocity.

MIM can be regarded as part of the scheme, and it is indispensable for low-Mach number flows. Therefore, determining the function of MIM in the shock instability and how to balance the existence of MIM between shock and incompressible flows is interesting. This question motivates this study to understand the MIM mechanism in shock instability and to propose a curing method to reach the aim of decreasing rather than increasing numerical dissipation.

The rest of this paper is organized as follows. Chapter 2 provides the governing equations and the Roe scheme. Chapter 3 discusses the effect of MIM on shock instability and proposes an improvement. Chapter 4 validates the analysis and improvement through numerical tests, especially through the odd–even decoupling test. Chapter 5 concludes the paper.

## 2. Governing Equations and the Roe Scheme

### 2.1 Governing Equations

The governing three-dimensional Euler equations can be written as follows:

$$\frac{\partial \boldsymbol{Q}}{\partial t} + \frac{\partial \boldsymbol{F}}{\partial x} + \frac{\partial \boldsymbol{G}}{\partial y} + \frac{\partial \boldsymbol{H}}{\partial z} = 0, \quad (1)$$

where $\boldsymbol{Q} = \begin{bmatrix} \rho \\ \rho u \\ \rho v \\ \rho w \\ \rho E \end{bmatrix}$ is the vector of conservation variables; $\boldsymbol{F} = \begin{bmatrix} \rho u \\ \rho u^2 + p \\ \rho uv \\ \rho uw \\ \rho uH \end{bmatrix}$,

$$G = \begin{bmatrix} \rho v \\ \rho uv \\ \rho v^2 + p \\ \rho vw \\ \rho vH \end{bmatrix}, \quad H = \begin{bmatrix} \rho w \\ \rho uw \\ \rho vw \\ \rho w^2 + p \\ \rho wH \end{bmatrix}$$ are the vectors of Euler fluxes; $\rho$ is the fluid density; $p$ is the pressure; $E$ is the total energy; $H$ is the total enthalpy; and $u, v, w$ are the velocity components in the Cartesian coordinates $(x, y, z)$, respectively.

## 2.2 Roe Scheme

The classical Roe scheme can be expressed as the following general sum form of a central term and a numerical dissipation term:

$$\tilde{F} = \tilde{F}_c + \tilde{F}_d, \tag{2}$$

where $\tilde{F}_c$ is the central term, and $\tilde{F}_d$ is the numerical dissipation term. For a cell face of the finite volume method,

$$\tilde{F}_{c,1/2} = \frac{1}{2}\left(\bar{F}_L + \bar{F}_R\right), \tag{3}$$

$$\bar{F} = U \begin{bmatrix} \rho \\ \rho u \\ \rho v \\ \rho w \\ \rho H \end{bmatrix} + \begin{bmatrix} 0 \\ n_x p \\ n_y p \\ n_z p \\ 0 \end{bmatrix}, \tag{4}$$

where $n_x$, $n_y$, and $n_z$ are the components of the face-normal vector, and $U = n_x u + n_y v + n_z w$ is the normal velocity on the cell face.

Following Ref. [2], a scale uniform framework for the shock-capturing scheme is proposed [23]. This framework is simple, has low computation cost, and is easy to be analyzed and improved.

$$\tilde{\mathbf{F}}_d = -\frac{1}{2}\left\{\xi\begin{bmatrix}\Delta\rho \\ \Delta(\rho u) \\ \Delta(\rho v) \\ \Delta(\rho w) \\ \Delta(\rho E)\end{bmatrix} + (\delta p_u + \delta p_p)\begin{bmatrix}0 \\ n_x \\ n_y \\ n_z \\ U\end{bmatrix} + (\delta U_u + \delta U_p)\begin{bmatrix}\rho \\ \rho u \\ \rho v \\ \rho w \\ \rho H\end{bmatrix}\right\}, \qquad (5)$$

where the five terms on the right side have explicit physical meanings. The first term $\xi$ is the basic upwind dissipation, which is just low-Mach Roe scheme [24]; the terms $\delta p_u$ and $\delta p_p$ are the velocity-difference-driven and pressure-difference-driven modifications for the cell face pressure, respectively; and the terms $\delta U_u$ and $\delta U_p$ are the velocity-difference-driven and pressure-difference-driven modifications for the cell face velocity, respectively.

For the Roe scheme,

$$\xi = \lambda_1, \qquad (6)$$

$$\delta p_u = \left(\frac{\lambda_5 + \lambda_4}{2} - \lambda_1\right)\rho\Delta U, \qquad (7)$$

$$\delta p_p = \frac{\lambda_5 - \lambda_4}{2}\frac{\Delta p}{c}, \qquad (8)$$

$$\delta U_u = \frac{\lambda_5 - \lambda_4}{2}\frac{\Delta U}{c}, \qquad (9)$$

$$\delta U_p = \left(\frac{\lambda_5 + \lambda_4}{2} - \lambda_1\right)\frac{\Delta p}{\rho c^2}, \qquad (10)$$

where $c$ is the sound speed, and the eigenvalues of the system are defined as follows:

$$\lambda_1 = \lambda_2 = \lambda_3 = |U|, \quad \lambda_4 = |U-c|, \quad \lambda_5 = |U+c|. \qquad (11)$$

Eqs. (6)–(11) are equivalent to the original form of the Roe scheme with the Roe average [23], which makes the following assumption reasonable from numerical viewpoint [2]:

$$\Delta(\rho\phi) = \rho\Delta\phi + \phi\Delta\rho, \qquad (12)$$

where $\phi$ represents one of the fluid variables.

Without the Entropy fix, Eqs. (6)–(11) can also be further simplified for easier analysis as follows:

$$\delta p_u = \max(0, c - |U|)\rho\Delta U, \qquad (13)$$

$$\delta p_p = \text{sign}(U)\min(|U|, c)\frac{\Delta p}{c}, \qquad (14)$$

$$\delta U_u = \text{sign}(U)\min(|U|, c)\frac{\Delta U}{c}, \tag{15}$$

$$\delta U_p = \max(0, c - |U|)\frac{\Delta p}{\rho c^2}. \tag{16}$$

## 3. Analysis of the MIM Role in Shock and Improvement

The $\delta U_p$ term is discussed as follows. When $|U| \to 0$, Eq. (16) becomes

$$\delta U_p = \frac{\Delta p}{\rho c}, \tag{17}$$

and can produce results with a weak checkerboard [4]. The reasonable interval of the order of coefficient of $\Delta p$ in $\delta U_p$ is $\left[c^{-1}, c^0\right]$. By decreasing the coefficient to lower than $c^{-1}$, the computational instability occurs because of the uncontrolled pressure checkerboard [4].

On the basis of the key idea of MIM, Ref. [18] derived the time-marching MIM, in which a two-dimensional steady form can be expressed as follows:

$$\delta U_p = \Delta t \left\{ n_x \left[ \frac{\left(\frac{\partial p}{\partial x}\right)_i}{2\rho_i} + \frac{\left(\frac{\partial p}{\partial x}\right)_{i+1}}{2\rho_{i+1}} - \frac{\left(\frac{\partial p}{\partial x}\right)_{i+1/2}}{\rho_{i+1/2}} \right] + n_y \left[ \frac{\left(\frac{\partial p}{\partial y}\right)_i}{2\rho_i} + \frac{\left(\frac{\partial p}{\partial y}\right)_{i+1}}{2\rho_{i+1}} - \frac{\left(\frac{\partial p}{\partial y}\right)_{i+1/2}}{\rho_{i+1/2}} \right] \right\}^{n-1}. \tag{18}$$

Eq. (18) introduces a third pressure-derivative term into the cell face velocity.

Compared with Eqs. (17) and (18), two equations have the same position in the scheme and similar mechanism of pressure difference and effect on suppressing the checkerboard problem. Therefore, Eq. (17) can be regarded as a version of MIM, which is indispensable for low-Mach number flows with the collocated-grid method.

However, for high-Mach number compressible flows, the decoupling of pressure and velocity disappears, and MIM becomes unnecessary. More importantly, MIM is unsuitable for compressible computation, as analyzed as follows.

From the perspective of the results, MIM suppresses the checkerboard problem, i.e., the pressure jump, and makes the pressure field smooth. However, shock is also a kind of pressure jump. Therefore, shock may be smoothed and destroyed by MIM.

From the perspective of mechanism, the main idea of MIM is to obtain cell face velocity by calculating the pressure gradient and interpolation of other terms. Then, the calculated value and the interpolation of pressure difference have matching errors for cell face velocity. The error is insignificant from the perspective of amount of numerical dissipation and is helpful for pressure and velocity coupling when the flow field is linear. However, the error may become too large to produce physical results for the nonlinear flow. Further, the staggered-grid method, which is another popular method for curing the checkerboard problem, cannot be adopted to capture shock because this method saves calculated pressure and velocity in different positions and cannot avoid the matching errors.

From the perspective of the Roe scheme itself, the Roe scheme holds the MIM for low-Mach number flows, and tends to automatically remove the MIM for high-Mach number flows. For the supersonic one-dimensional flows, the $\delta U_p$ term is equal to zero according to Eq. (16). However, given that the Roe scheme is developed from a one-dimensional Riemann problem and it is simply extended for multi-dimensional flows, the MIM that should disappear survives on the grid face approximately parallel to the flow direction, as shown in Fig. 1 (a). In this figure, $|U| \to 0$, although the flow speed $V$ may be supersonic.

Notably, cells of low Mach number may exist in shock, as shown in Fig. 1 (b). Flows of such cells are strongly nonlinear and should be treated as shock rather than as incompressible flows, although their Mach numbers are low. The MIM should then disappear. For this condition, the traditional judging criterion of Mach number becomes invalid, and a shock detector is needed to identify this condition.

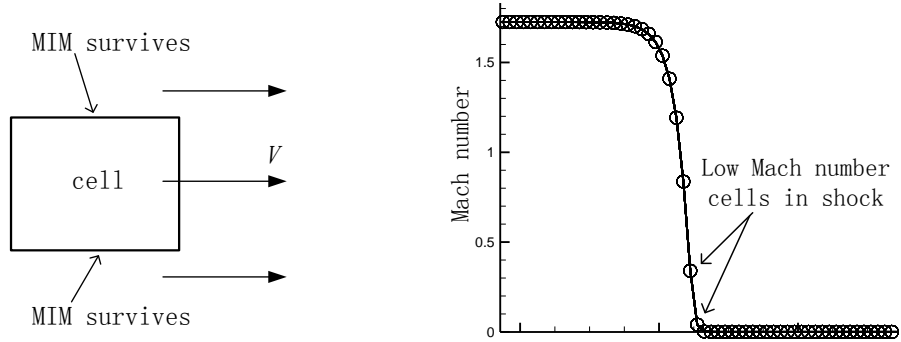

(a) Cell face parallel to the flow direction    (b) Low Mach number cell in shock

Fig. 1 Conditions of the unexpected existence of the MIM

According to the above discussions, two rules can be proposed to avoid harmful MIM:

(1) For high Mach number flows, $\delta U_p = 0$ under any circumstances;

(2) For low Mach number flows in shock, $\delta U_p$ should also be equal to zero.

For the first rule, a coefficient for $\delta U_p$ can be designed as follows:

$$s_1 = 1 - f^8(M), \tag{19}$$

where the Mach number $M = \dfrac{\sqrt{u^2 + v^2 + w^2}}{c}$, and the function $f$ is proposed [25, 26] as follows:

$$f(\varphi) = \min\left(\varphi \dfrac{\sqrt{4 + (1-\varphi^2)^2}}{1+\varphi^2}, 1\right), \tag{20}$$

where $\varphi$ represents any of the variables. As shown in Fig. 2, the function $f^8$ remains nearly zero when Mach number is lower than 0.3, and has a smooth transition near the sound speed. However, $s_1$ is based on the Mach number and cannot satisfy the second rule. Therefore, another coefficient is proposed as follows:

$$s_2 = f^8(b), \qquad (21)$$

where $b$ is a shock detector. A simple and effective design of $b$ [8, 12], which searches pressure jumps among the cell face and its neighbor faces, is shown as the following two-dimensional form:

$$b_{i+1/2,j} = \min\left(P_{i+1/2,j}, P_{i+1,j-1/2}, P_{i+1,j+1/2}, P_{i,j-1/2}, P_{i,j+1/2}\right), \qquad (22)$$

$$P_{i+1/2,j} = \min\left(\frac{p_{i,j}}{p_{i+1,j}}, \frac{p_{i+1,j}}{p_{i,j}}\right). \qquad (23)$$

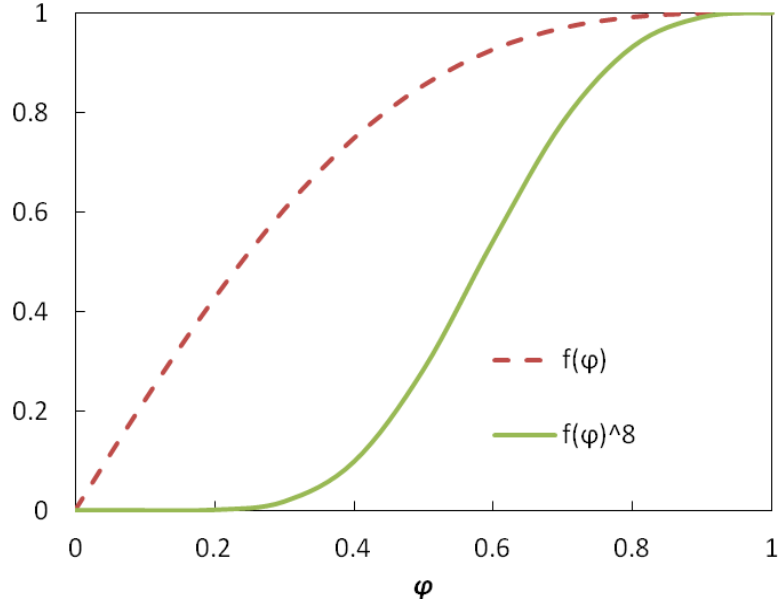

Fig. 2 Functions of $f$ and $f^8$

Therefore, the MIM term $\delta U_p$ Eq. (16) can be improved by multiplying the coefficients $s_1$ and $s_2$.

$$\delta U_p = s_1 s_2 \max(0, c - |U|)\frac{\Delta p}{\rho c^2}. \qquad (24)$$

According to preceding discussion, the coefficient $s_1$ is based on physical variable Mach number, and the shock-detector coefficient $s_2$ is based on numerical shock thickness and number of low Mach number nodes in shock. Then, $s_1$ is indispensable, but $s_2$ maybe replaced by other better methods such as high-order reconstruction, which can effectively reduce the number of low Mach nodes in shock. Therefore, the following modification with only $s_1$ is also considered:

$$\delta U_p = s_1 \max\left(0, c - |U|\right) \frac{\Delta p}{\rho c^2}. \tag{25}$$

For the improved Roe scheme with Eq. (24) or (25), the numerical dissipation does not increases and sometimes even decreases for weakening the MIM unlike the other curing methods. Therefore, the improved Roe scheme provides a way to cure shock instability by decreasing numerical dissipation.

In addition, according to the two rules in Chapter 3, the $\delta U_p$ term can be directly set to zero for supersonic flows:

$$\delta U_p = 0. \tag{26}$$

Eq. (26) can be used to test the ultimate effect of MIM on shock instability and to validate the improvement of Eq. (24) or (25).

## 4. Numerical Tests

### 4.1 Low Mach Number Inviscid Flows Around a Cylinder

The two-dimensional Euler flow past a cylinder is a typical test case. The computation is performed with an inflow Mach number of 0.01 and the 100*72 O-type grid points along the radius and circumference, respectively.

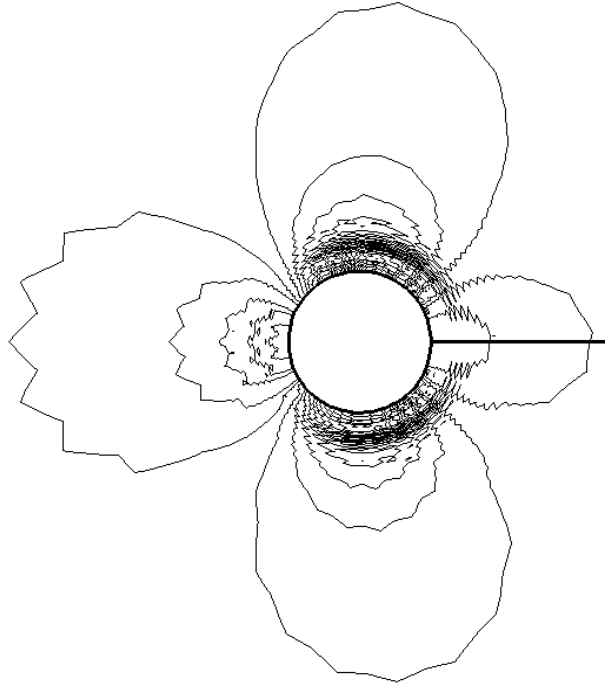

Fig. 3 Pressure contours by the all-speed Roe scheme with Eq. (26)  $\delta U_p = 0$

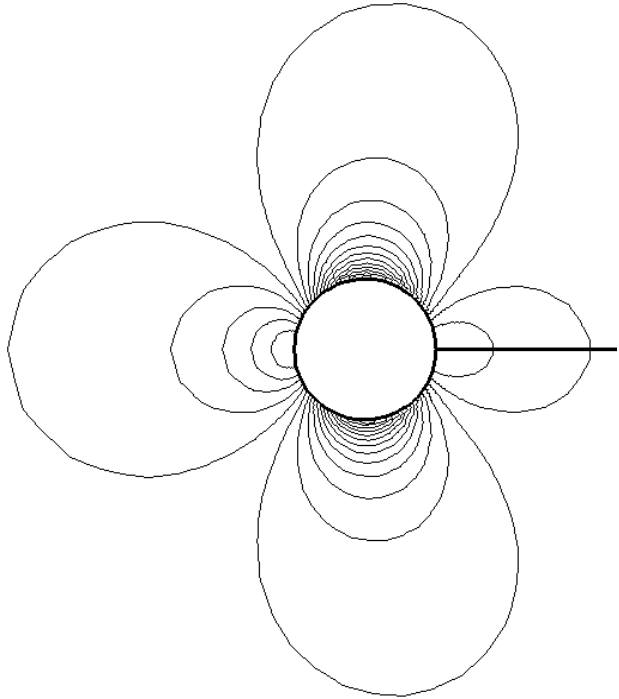

Fig. 4 Pressure contours by the all-speed Roe scheme with Eq. (16) or (24) or (25)

Adopting an all-speed Roe scheme [4] with Eq. (29) $\delta U_p = 0$, i.e., MIM is completely removed, a severe pressure checkerboard quickly appears as shown in Fig. 3, even if the good solution in Fig. 4 is used as the initial field. The effect increases over time, leading to instability. When Eq. (29) is replaced by Eq. (16) or (24) or (25), the checkerboard is suppressed by MIM and the same convergence solutions can be obtained as shown in Fig. 4. The results also indicate that the improvement of Eq. (24) or (25) does not affect MIM for low Mach number flows.

The checkerboard means pressure zigzag-shape jump as shown in Fig. 3, and can be regarded as some kind of pseudo-shock. Now that MIM can smooth this pseudo-shock, can MIM smooth physical shock? The preceding discussions and following numerical cases give certain answer.

### 4.2 Odd–Even Decoupling Test

### 4.2.1 Computational Object and Method

This test case was designed by Quirk [5]. A planar shock moves in a duct where a centerline grid is odd–even perturbed, as shown in Eq. (27).

$$Y_{i,j,\text{mid}} = \begin{cases} Y_{j,\text{mid}} + \varepsilon_y \Delta Y, & \text{for } i \text{ even,} \\ Y_{j,\text{mid}} - \varepsilon_y \Delta Y, & \text{for } i \text{ odd.} \end{cases} \quad (27)$$

where a larger value of $\varepsilon_y$ results in a more serious odd–even decoupling. This test case is important because any scheme that does not survive it meets more or less shock instability problems in other classical cases.

The initial conditions are given as $(\rho, p, u, v)_L = \left(\dfrac{1512}{205}, \dfrac{251}{6}, \dfrac{175}{36}, 0\right)$ and $(\rho, p, u, v)_R = (1.4, 1, 0, 0)$, which produce a normal shock with a moving Mach number of 6. The computational mesh has 20*800 cells in the Y and X directions, with $\Delta Y = 1$ and $\Delta X = 1$. The schemes are adopted with first-order accuracy unless otherwise specified. Apart from the original Roe scheme and its improvement in Eq. (24) or (25) or (26), following condition Eqs. (28)–(29) is also considered.

The following entropy fix for Eqs. (28)–(29) is also adopted for comparison because it is commonly used.

$$\lambda_i = \begin{cases} \lambda_i, & \lambda_i \geq h, \\ \dfrac{1}{2}\left(\dfrac{\lambda_i^2}{h} + h\right), & \lambda_i < h, \end{cases} \tag{28}$$

$$h = \varepsilon_\lambda \max(\lambda_i), \tag{29}$$

where $\varepsilon_\lambda$ is a constant with a commonly adopted value of 0.05 to 0.2. When the value of $\varepsilon_\lambda$ is zero, no entropy fix is used with the Roe scheme.

Given that the increase in $\varepsilon_y$ can amplify the odd–even decoupling phenomenon and can bring more difficulty in computation, the two conditions of $\varepsilon_y = 10^{-4}$ and $\varepsilon_y = 0.1$ are investigated as follows.

### 4.2.2 Condition of $\varepsilon_y = 10^{-4}$

When $\varepsilon_y = 10^{-4}$, the unexpected MIM is activated and the Roe scheme without the entropy fix fails. The shock is deformed from middle part at $X \approx 220$, as shown in Fig. 5 (a), and is completely destroyed, as shown in Fig. 5 (b).

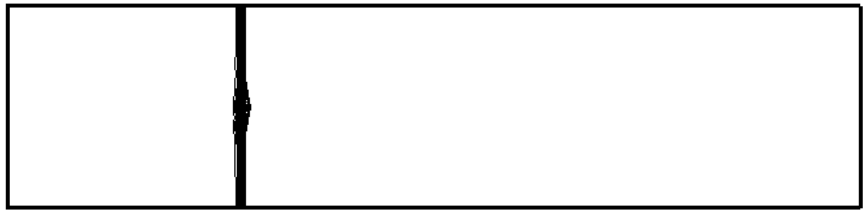

(a) Initial shock deformation

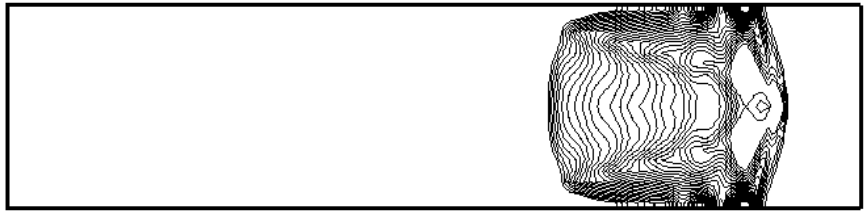

(b) Complete smoothening of shock

Fig. 5 Roe scheme without the entropy fix $\varepsilon_\lambda = 0$

The result of adopting the entropy fix only achieves slight improvement with a common value of 0.05 for $\varepsilon_\lambda$, as shown in Fig. 6 (a). By increasing the value of $\varepsilon_\lambda = 0.05$ up to 0.2, the shock recovers to normal, as shown in Fig. 6 (b).

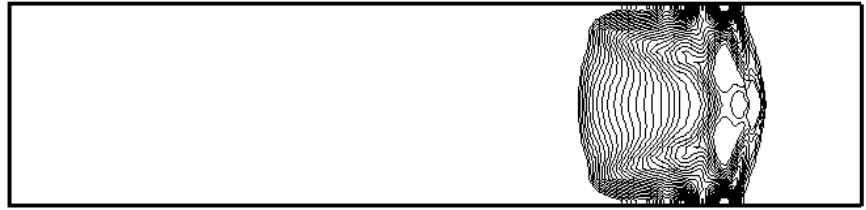

(a) $\varepsilon_\lambda = 0.05$

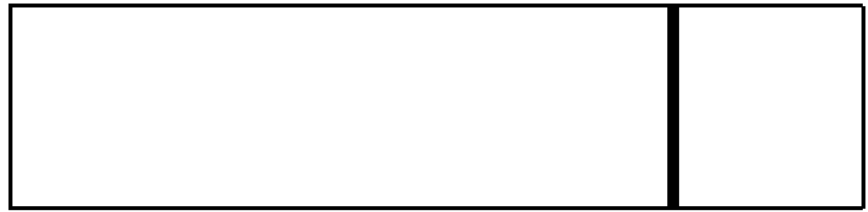

(b) $\varepsilon_\lambda = 0.2$

Fig. 6 Roe scheme with the entropy fix

By decreasing the numerical dissipation of the MIM, the Roe scheme can obtain improved results without adding any extra dissipation, as shown in Fig. 7. Adopting Eq. (26), the solution becomes correct as shown in Fig. 7 (a), which indicates that MIM does play a very important role for shock instability. As shown in Fig. 7 (b), Eq. (24) produces the same solution as Eq. (26), which validates that they are almost equivalent.

Eq. (25) removes the shock detector $s_2$ from the improvement, and thus a sight shock deformation occurs as shown in Fig. 7 (c). This result indicates that cells of low Mach number in shock do have obvious effect on MIM and shock instability.

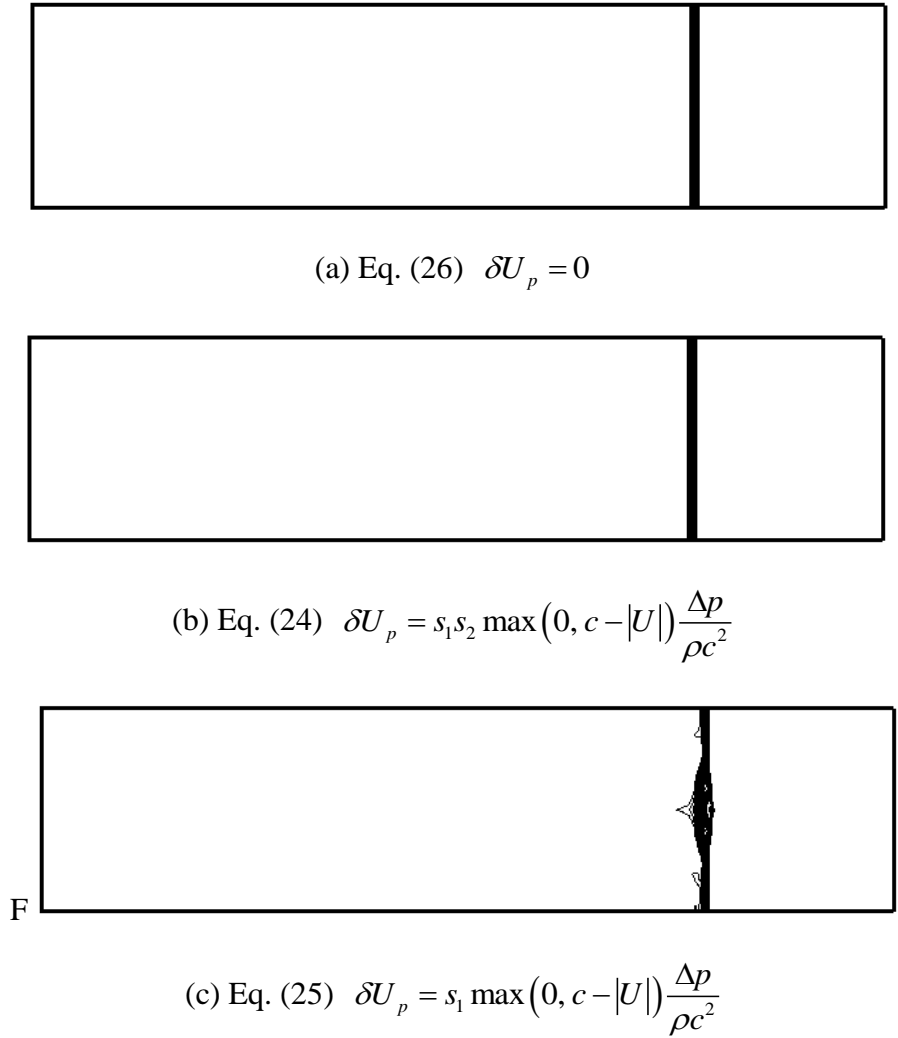

(a) Eq. (26) $\delta U_p = 0$

(b) Eq. (24) $\delta U_p = s_1 s_2 \max(0, c - |U|) \dfrac{\Delta p}{\rho c^2}$

(c) Eq. (25) $\delta U_p = s_1 \max(0, c - |U|) \dfrac{\Delta p}{\rho c^2}$

Fig. 7 Improved Roe schemes with modified $\delta U_p$

For practical computation, the high-order accuracy methods of space reconstruction are usually employed such as MUSCL-TVD (the monotone upstream-centered schemes for conservation laws) [27, 28], WENO (the weighted essentially nonoscillatory scheme) [29, 30], and DG (the discontinuous Galerkin) methods [31–33]. Therefore, the compatibility of preceding improvement and high-order reconstruction should be discussed. In this paper, second-order accuracy MUSCL [27] reconstruction with minmod limiter for TVD characteristic is adopted because it is widely used.

As shown in Fig. 8 (a) and (b), MUSCL-TVD reconstruction keeps the advantages of Eqs. (26) and (24), and make the shock shape by Eq. (25) better in Fig. 8 (c). The reason may be due to the much fewer nodes in shock. Compared with thick shock shape in Fig. 7 (a) and approximately 10 nodes in shock for the first scheme in Fig. 1 (b), the shock becomes thin in Fig. 8 (a) and only three nodes are needed to form the shock in Fig. 8 (d), where number of low Mach node is only one. Clearly, fewer nodes cause the lesser possibility of inducing shock instability. Therefore, high-order reconstruction can replace the coefficient $s_2$ to a great degree.

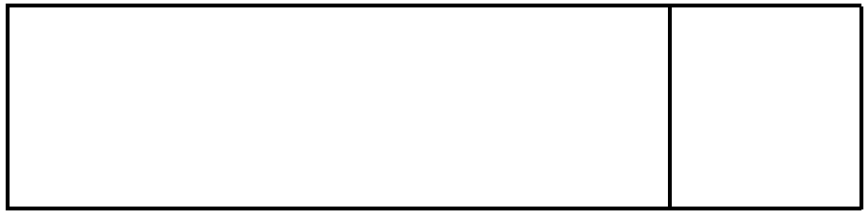

(a) Eq. (26) $\delta U_p = 0$

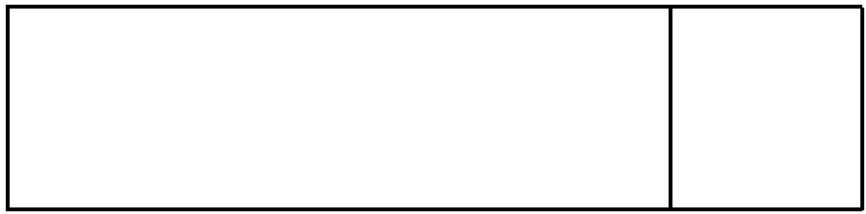

(b) Eq. (24) $\delta U_p = s_1 s_2 \max\left(0, c - |U|\right) \dfrac{\Delta p}{\rho c^2}$

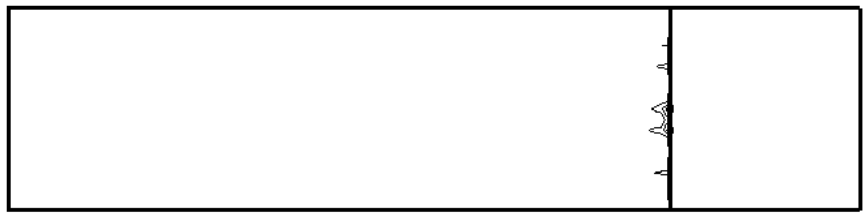

(c) Eq. (25) $\delta U_p = s_1 \max\left(0, c - |U|\right) \dfrac{\Delta p}{\rho c^2}$

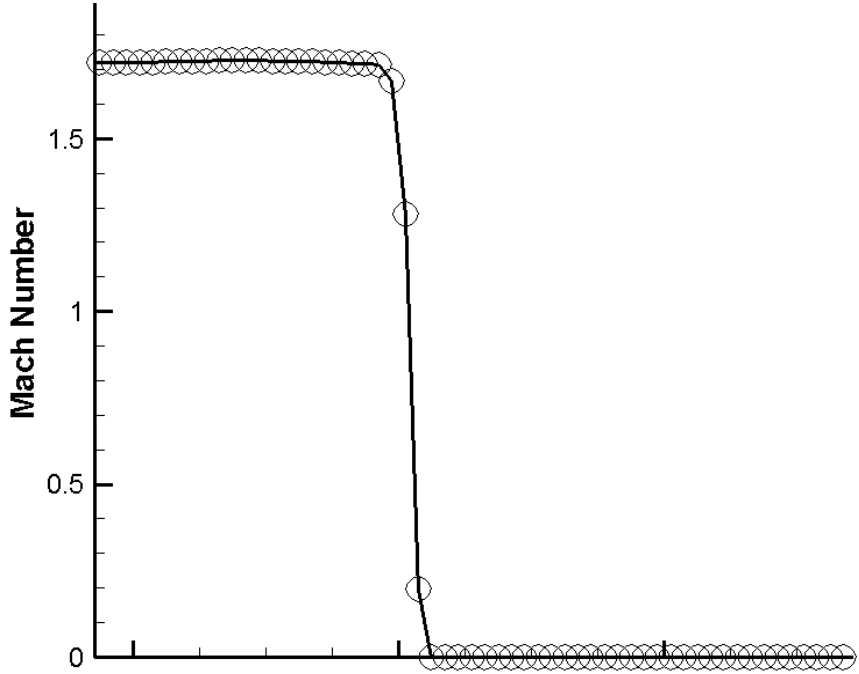

(d) Mach number distribution by Eq. (26) $\delta U_p = 0$

Fig. 8 Improved Roe schemes with MUSCL-TVD reconstruction

### 4.2.3 Condition of $\varepsilon_y = 0.1$

The computation becomes more difficult when the value of $\varepsilon_y$ is increased to 0.1. As shown in Fig. 9, the shock is also seriously deformed even when $\varepsilon_\lambda = 0.2$. By removing the MIM completely by Eq. (26) $\delta U_p = 0$, the shock is also twisted to a great extent, as shown in Fig. 10 (a). This finding indicates that the MIM is not the only factor for shock instability. Even so, the MIM can be regarded as the most important factor because removing it is clearly better than adding the large entropy fix. The result in Fig. 10 (b) using the improved Roe scheme with Eq. (24) is similar to that in Fig. 10 (a), thus further validating Eq. (24). Some differences between Fig. 10 (a) and (b) may be due to a few activations of the MIM in a smoothed area of shock, thus making the shock detector invalid. When the shock detector is completely removed, the shock deformation in Fig. 10 (c) becomes much more serious than that in Fig. 10 (b). It further validates the importance of low Mach number cells in shock.

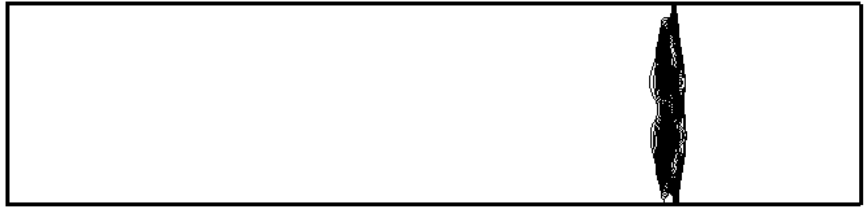

Fig. 9 Roe scheme with the entropy fix $\varepsilon_\lambda = 0.2$

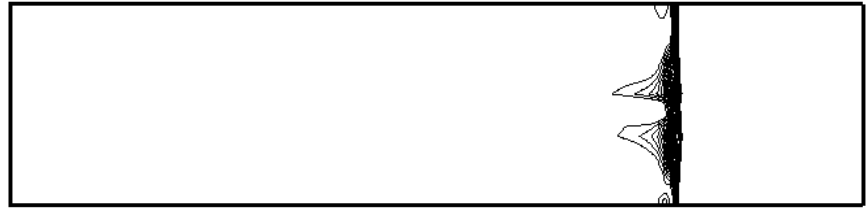

(a) Eq. (26) $\delta U_p = 0$

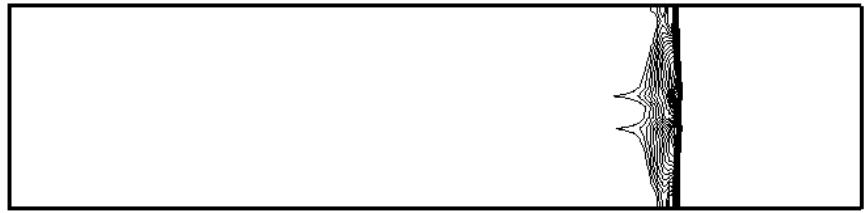

(b) Eq. (24) $\delta U_p = s_1 s_2 \max\left(0, c - |U|\right) \dfrac{\Delta p}{\rho c^2}$

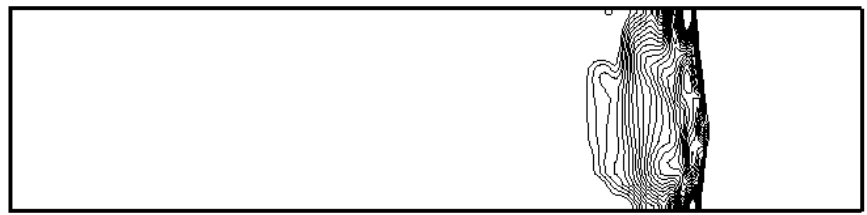

(c) Eq. (25) $\delta U_p = s_1 \max\left(0, c - |U|\right) \dfrac{\Delta p}{\rho c^2}$

Fig. 10 Improved Roe schemes with modified $\delta U_p$

As shown in Fig. 11, the improved Roe schemes with MUSCL-TVD reconstruction also produce good results similar to those in Fig. 8. Even the shock deformation is serious for the first order accuracy in Fig. 10, for high-order accuracy Eq. (26) and Eq. (24) produce almost perfect results in Fig. 11 (a) and (b), and Eq. (25) gives acceptable solution in Fig. 11 (c).

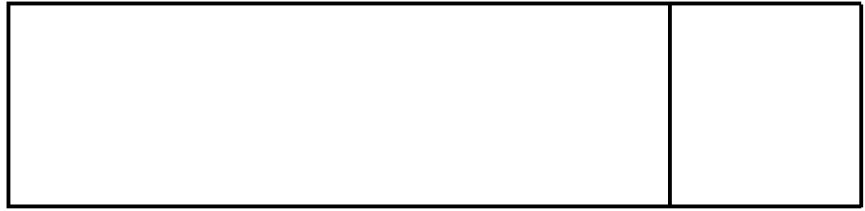

(a) Eq. (26) $\delta U_p = 0$

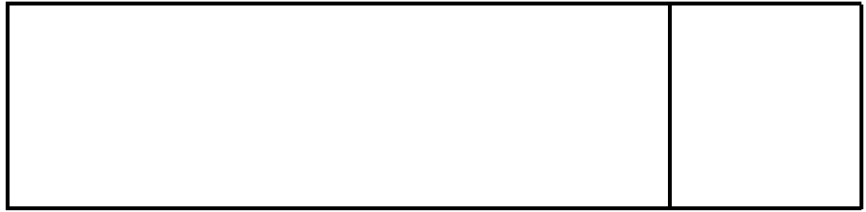

(b) Eq. (24) $\delta U_p = s_1 s_2 \max\left(0, c - |U|\right) \dfrac{\Delta p}{\rho c^2}$

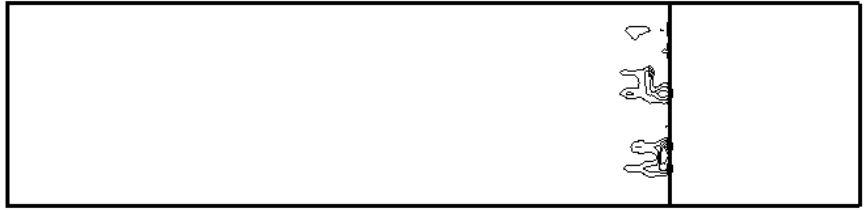

(c) Eq. (25) $\delta U_p = s_1 \max\left(0, c - |U|\right) \dfrac{\Delta p}{\rho c^2}$

Fig. 11 Improved Roe schemes with MUSCL-TVD reconstruction

**4.3 Kinked Mach Stem**

The kinked Mach stem is another popular shock instability phenomenon that occurs when an inclined moving shock wave is reflected from a wall to form a double-Mach reflection. The shock is initially set up to be inclined at an angle of 60°, and it has a Mach number of 10. In Fig. 12, the density contours are shown on a 200*800 mesh at $t$=0.2.

Fig. 12 gives expected results. For the classical Roe scheme without the entropy fix, the Mach stem is severely kinked so that a non-physical triple point called "the kinked Mach stem" appears, as shown in Fig. 12 (a). With the entropy fix, the kinked Mach stem is improved but also becomes obvious, as shown in Fig. 12 (b). By adopting the improved Roe scheme with removing the MIM completely by Eq. (26), the results shown in Fig. 12 (c) are much better than those by the entropy fix, and the triple point appears to be negligible. It proves once again that the MIM is the most important factor for shock instability, although the MIM is not the only factor. The improved Roe scheme with Eq. (24) produces almost the same results in Fig. 12 (d) as Fig. 12 (c). It indicates that Eq. (24) has ability of removing the MIM completely as Eq. (26) for shock. Adopting Eq. (25) in Fig. 12 (e), the kinked Mach stem problem becomes slightly more obvious compared with Fig. 12 (d). When the MUSCL-TVD reconstruction is also adopted, however, the kinked Mach stem problem is completely cured by Eq. (24) and (25) as shown in Fig. 12 (f) and Fig. 12 (g), respectively. Therefore, the same conclusions can be obtained for different problems of odd–even decoupling and kinked Mach stem.

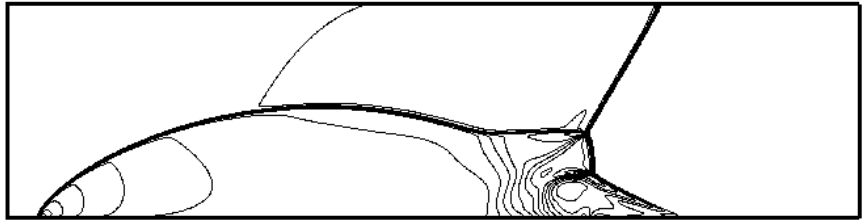

(a) Roe scheme without the entropy fix $\varepsilon_\lambda = 0$

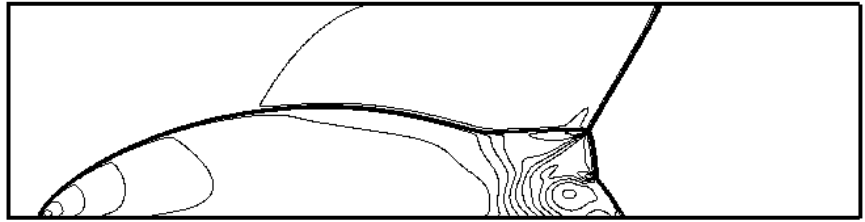

(b) Roe scheme with the entropy fix $\varepsilon_\lambda = 0.2$

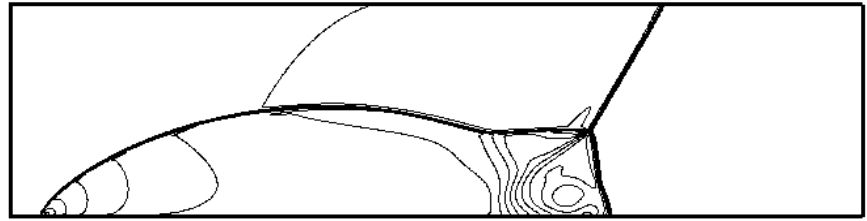

(c) Improved Roe scheme with the Eq. (26)

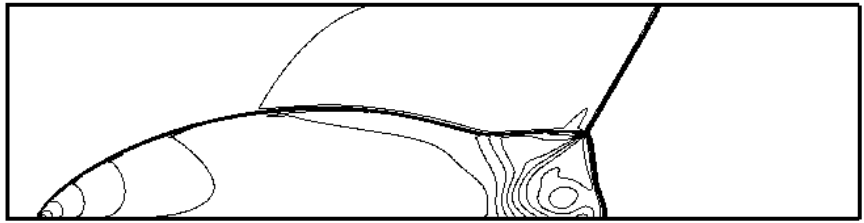

(d) Improved Roe scheme with the Eq. (24)

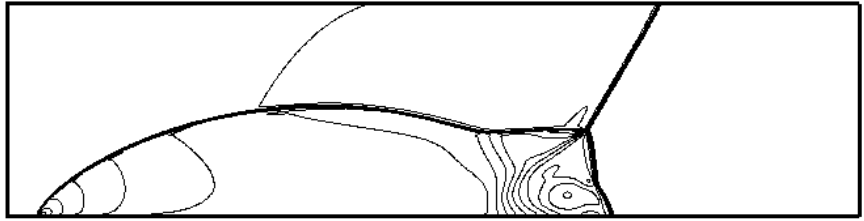

(e) Improved Roe scheme with the Eq. (25)

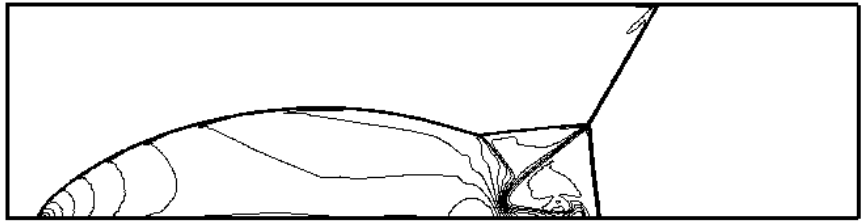

(f) Improved Roe scheme with Eq. (24) and MUSCL-TVD reconstruction

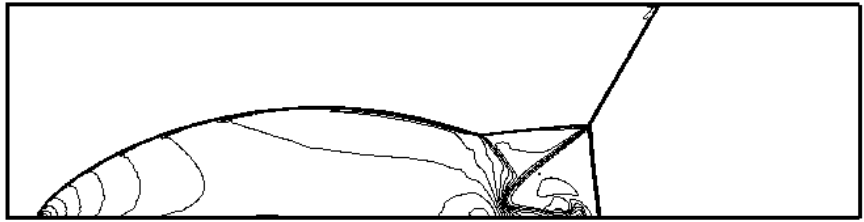

(g) Improved Roe scheme with Eq. (25) and MUSCL-TVD reconstruction

Fig. 12 Density contours of the double-Mach reflection

**4.4 Carbuncle Phenomenon of Supersonic Flow around a Circular Cylinder**

The supersonic flow around a circular cylinder is also a well-known test case to examine the catastrophic carbuncle failings of a few schemes. This test is computed with the free stream Mach number of 20 and the first-order accuracy on a grid with 20*160 cells in the radial and circumferential directions. The so-called carbuncle phenomenon is shown in Figs. 13 (a) and (b), where the entropy fix cannot offer help because the Mach number is too high. On the contrary, the results of the improved Roe scheme with Eq. (24) unsurprisingly do not exhibit this kind of shock instability, as shown in Fig. 13 (c). The scheme with Eq. (25) produces almost the same results in Fig. 13 (d) as Fig. 13 (c), because in this case there is no low Mach number node in shock as shown in Fig. 14. Therefore, Eq. (25) is just equal to Eq. (24).

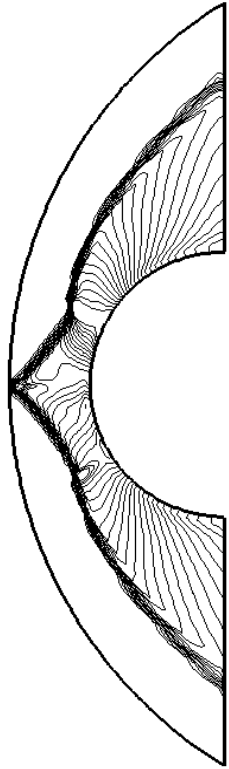 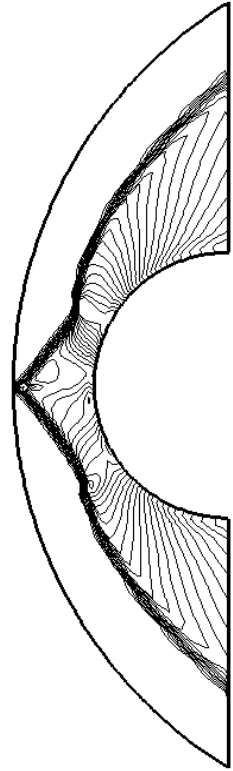

(a) Roe scheme with $\varepsilon_\lambda = 0$      (b) Roe scheme with $\varepsilon_\lambda = 0.2$

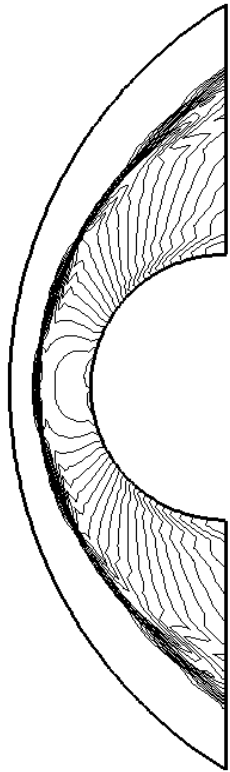 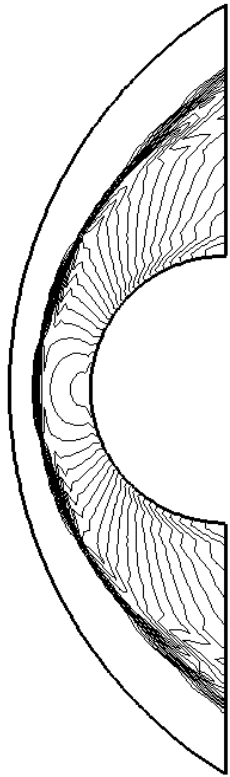

(c) Improved Roe scheme with the Eq. (24) (d) Improved Roe scheme with the Eq. (25)

Fig. 13 Pressure contours of the supersonic flow ($M_\infty = 20$) around a circular cylinder

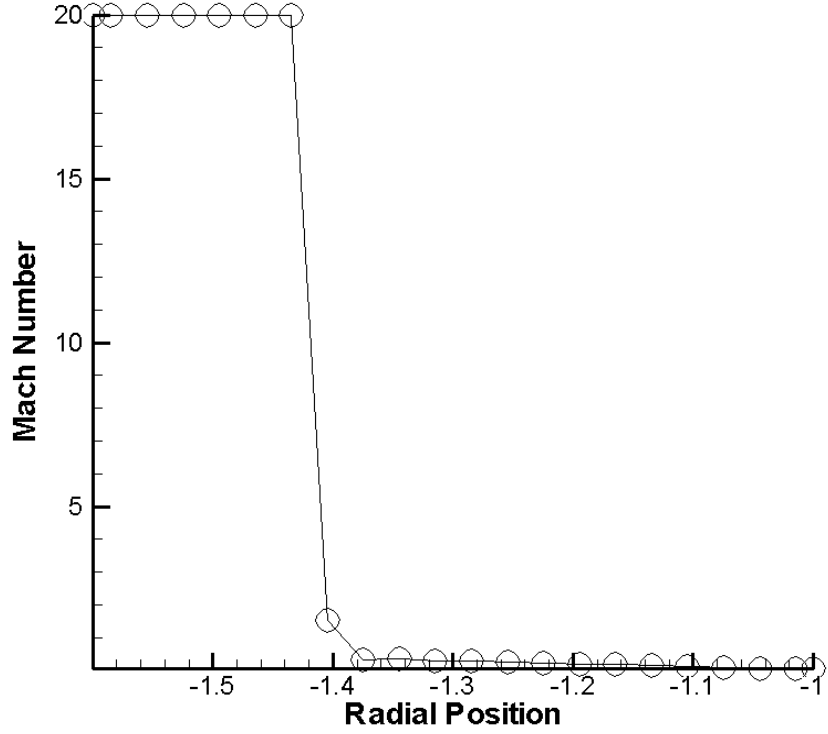

Fig. 14 Mach number distribution along centre radial line by Eq. (25)

## 5. Conclusions

The numerical dissipation term $\delta U_p$ of the Roe scheme is considered a version of the MIM, and the role of MIM is investigated for the shock instability problem. This study finds that the MIM is the most important factor for the shock instability, although it is not the only factor. The MIM is indispensable for low-Mach number flows to suppress the pressure checkerboard, but leads to shock instability because of the unexpected activation on the cell face parallel to the flow when the Mach number is high or low but in shock. Therefore, the following three rules should be satisfied:

(1) For high Mach number flows, $\delta U_p = 0$;

(2) For low Mach number flows in shock, $\delta U_p = 0$;

(3) For low Mach number flows, $\delta U_p$ of the Roe scheme should be kept.

According to the rules, two coefficients are proposed on the basis of the Mach number and a shock detector. By multiplying $\delta U_p$ by two coefficients, above all three rules can be satisfied. The function of the shock-detector coefficient is important, but can be partly replaced by the high order reconstruction. Therefore, for general practical computation, the acceptable result can be obtained with only Mach number coefficient, which is very simple and low computational cost.

Through validation of classical numerical cases including low Mach number flows and three shock instability phenomena, odd–even decoupling, kinked Mach stem and carbuncle, the improved Roe scheme based on proposed coefficients can achieve the aim of decreasing numerical dissipation to cure shock instability, and take into account requirement of computation of low Mach number flows.

**Acknowledgments**

This work is supported by Project 51276092 of the National Natural Science Foundation of China.